\input{aipcheck}

\documentclass[
    ,final            
  ]
  {aipproc}

\layoutstyle{6x9}

\begin{document}

\phantom{hola}

\vspace{-1.4cm}

\title{The oblique S parameter in higgsless electroweak models\footnote{Talk given at {\it QCD@work 2012, International Workshop on Quantum Chromodynamics: Theory and Experiment}, 18-21th June (2012), Lecce
(Italy). IFIC/12-64 report}}


\author{Ignasi Rosell}{
  address={Departamento de Ciencias F\'\i sicas, Matem\'aticas y de la Computaci\'on, Escuela Superior de Ense\~nanzas T\'ecnicas ESET, Universidad CEU Cardenal Herrera, c/ Sant Bartomeu 55, \\ E-46115 Alfara del Patriarca, Val\`encia, Spain \\ \phantom{hi}}, 
  altaddress={IFIC, Universitat de Val\`encia -- CSIC, Apt. Correus 22085, E-46071 Val\`encia, Spain}
}

\begin{abstract}
 We present a one-loop calculation of the oblique S parameter within Higgsless models of electroweak symmetry breaking. We have used a general effective Lagrangian with at most two derivatives, implementing the chiral symmetry breaking $SU(2)_L\otimes SU(2)_R\to SU(2)_{L+R}$ with Goldstones, gauge bosons and one multiplet of vector and axial-vector resonances. 
The estimation is based on the short-distance constraints and the dispersive approach proposed by Peskin and Takeuchi.
\end{abstract}

\maketitle

\section{Introduction}

The Standard Model (SM) provides an extremely successful description of the electroweak and strong interactions. A key feature of this theoretical framework is the particular mechanism adopted to break the electroweak gauge symmetry $SU(2)_L\times U(1)_Y$ to the electromagnetic subgroup $U(1)_{\mathrm{QED}}$, so that the $W$ and $Z$ bosons become massive~\cite{HK-mechanism}. 

The SM implements the Electroweak Symmetry Breaking (EWSB) through an $SU(2)_L$ doublet of complex scalars $\Phi(x)$ and a potential with non-trivial minima. The vacuum expectation value of the scalar doublet generates the needed spontaneous symmetry breaking, giving rise to three Goldstone bosons which, in the unitary gauge, become the longitudinal polarizations of the gauge bosons. Since $\Phi(x)$ contains four real fields, one massive neutral scalar survives in the physical spectrum: the Higgs boson. 

The LHC has just discovered a new particle around $125\,$GeV~\cite{LHC}, which could be the SM Higgs. If this particle is not the SM Higgs (it could be a non standard one or a resonance of spin 0 or 2), we should look for alternative mechanisms of mass generation, satisfying the many experimental constraints which the SM has successfully fulfilled so far. 

In the limit where the $U(1)_Y$ coupling $g'$ is neglected, the scalar sector of the SM Lagrangian is invariant under global $SU(2)_L\otimes SU(2)_R$ transformations. Taking now the limit of a heavy Higgs~\cite{AB:80}, one recovers the universal model-independent lowest-order Goldstone Lagrangian associated with the symmetry breaking $SU(2)_L\otimes SU(2)_R\to SU(2)_{L+R}$. In Quantum Chromodynamics (QCD) this Lagrangian describes the dynamics of pions at $\mathcal{O} (p^2)$ (two derivatives) in terms of $f_\pi$, the pion decay constant. The same Lagrangian with $f_\pi \to v = 246\:\mathrm{GeV}$ describes the Goldstone boson dynamics associated with the EWSB. 


In strongly-coupled models the gauge symmetry is dynamically broken by means of some non-perturbative interaction. Usually, theories of this kind do not contain any fundamental Higgs, bringing instead resonances of different types as happens in QCD~\cite{EWSB}. Technicolor~\cite{technicolor}, the most studied strongly-coupled model, introduces an asymptotically-free QCD replica at TeV energies which breaks the symmetry in a similar way as chiral symmetry in QCD, appearing a tower of heavy resonances.

In Ref.~\cite{paper} we have reanalyzed the oblique $S$ parameter~\cite{Peskin:92} on strongly-coupled models. We have performed a one-loop calculation of this important quantity within an effective theory including the electroweak Goldstones and resonance fields. The theoretical framework is completely analogous to the Resonance Chiral Theory (RChT) description of QCD at GeV energies~\cite{RChT}. We have made use of the procedure developed to compute the low-energy constants of Chiral Perturbation Theory (ChPT) at the next-to-leading order (NLO) through a matching with R$\chi$T~\cite{estimating-LECs,L10}. The estimation of $S$ in strongly-coupled electroweak models is equivalent to the calculation of $L_{10}$ in ChPT~\cite{L10}. 

Several one-loop estimates of the electroweak $S$ and $T$ parameters in Higgless models have appeared recently~\cite{other}. In this work, and following the dispersive approach suggested in~\cite{Peskin:92}, we have not used any cut-off and a more general Lagrangian has been considered. A crucial ingredient of this approach is the assumed high-energy behaviour of the relevant Green functions.

\section{The calculation}

In Ref.~\cite{paper} we have concentrated on the $S$ parameter, for which a useful dispersive representation was introduced by Peskin and Takeuchi\cite{Peskin:92}.
%
The convergence of this dispersion relation requires a vanishing spectral function at short distances. 

We have considered an effective framework containing the SM gauge bosons coupled to the electroweak Goldstones and the lightest vector and axial-vector resonances, which can induce sizeable corrections to $S$. We have only assumed the SM pattern of EWSB, {\it i.e.} the theory is symmetric under $SU(2)_L\otimes SU(2)_R$ and becomes spontaneously broken to the diagonal subgroup $SU(2)_{L+R}$. 
The Lagrangian can be found in Ref.~\cite{paper} and the two-particle spectral functions are determined in terms of seven resonance parameters.

The tree-level contributions to the gauge-boson vacuum polarization $\Pi_{30}(s)$ lead to the well-known leading-order (LO) result which determines the $S$ parameter at LO~\cite{Peskin:92}:
\begin{equation}
\left. \Pi_{30}(s) \right|_{\mathrm{LO}} = \frac{g^2  \tan{\theta_W} \,s}{4}  \left(\frac{v^2}{s}\!+\!  \frac{F_V^2}{M_V^2-s} \!-\! \frac{F_A^2}{M_A^2-s} \right)  , \qquad
S_{\mathrm{LO}} = 4\pi \left( \frac{F_V^2}{M_V^2}\! -\! \frac{F_A^2}{M_A^2} \right)  . 
\label{eq.TL}
\end{equation}

As it has been explained, the NLO contribution is most efficiently obtained through a dispersive calculation. We have considered two-particle cuts with two Goldstones or one Goldstone plus one massive resonance, either vector or axial-vector. The explicit results for the different spectral functions are given in Ref.~\cite{paper}. The total NLO result, including the tree-level exchanges, and the $S$ parameter can be written in the form~\cite{paper,L10}
\begin{equation}
\left. \Pi_{30}(s) \right|_{\mathrm{NLO}}   = \frac{g^2\tan{\theta_W}\,s }{4}   \!\left(\!\frac{v^2}{s}\!\!+\!\!  \frac{F_{V}^{r\,2}}{M_{V}^{r\,2}\!-\!s} \!\!-\!\! \frac{F_{A}^{r\,2}}{M_{A}^{r\,2}\!-\!s} \! \!+\!\! \overline{\Pi}(s)\!\right)\! , \,\,\,\,
S_{\mathrm{NLO}} = 4\pi \!\left(\! \frac{F_{V}^{r\,2}}{M_{V}^{r\,2}} \!\!-\!\! \frac{F_{A}^{r\,2}}{M_{A}^{r\,2}} \!\right) \! \!+\!\! \overline{S}  ,
\end{equation}
where $F_R^r$ and $M_R^r$ are ``renormalized'' couplings which properly define the resonance poles at the one-loop level. 


The number of unknown couplings can be reduced using short-distance information: 
\begin{enumerate}
\item {\bf Weinberg sum rules}. Since we have assumed that weak isospin and parity are good symmetries of the strong dynamics, the correlator $\Pi_{30}(s)$ can be written in terms of the vector and axial-vector two-point functions
~\cite{Peskin:92}.
%
Assuming the two Weinberg sum rules (WSRs)~\cite{WSR} and taking the results of Eq.~(\ref{eq.TL}), one gets $F_{V}^2 - F_{A}^2  = v^2 $ and $F_{V}^2  \,M_{V}^2 - F_{A}^2 \, M_{A}^2  = 0$. 
It is likely that the first of these sum rules is also true in gauge theories with non-trivial ultraviolet fixed points. However, the second WSR is questionable in some Technicolour scenarios. %
Therefore, if both WSRs are valid, $F_V$ and $F_A$ are determined at LO in terms of the resonance masses and $M_A > M_V$.
%
%

At NLO the computed spectral functions should behave also as dictated by this pattern. The first WSR provides 
three constraints. We have also analyzed the impact of considering the second WSR, a fourth constraint.

After imposing the short-distance conditions on the spectral function, one has to apply the same constraints to the real part of the correlator:
assuming the two WSRs it is possible to fix the couplings $F_V^r$ and $F_A^r$ up to NLO~\cite{paper,L10}.
%
%

\item {\bf  Vector form factor}. The two-Goldstone matrix element of the vector current defines the vector form factor (VFF). Imposing that the VFF vanishes at $s\rightarrow \infty$, one gets the LO constraint $F_V G_V  = v^2$~\cite{RChT}.

\item {\bf  Axial form factor}. The matrix element of the axial current between one Goldstone and one photon is parameterized by the axial form factor (AFF). Requiring the AFF to vanish at $s\to\infty$ implies that
$ F_V - 2 G_V  = F_A\left(2\kappa +\sigma\right)$~\cite{L10}.
\end{enumerate}

\section{Phenomenology}
\begin{figure}
\includegraphics[scale=0.8]{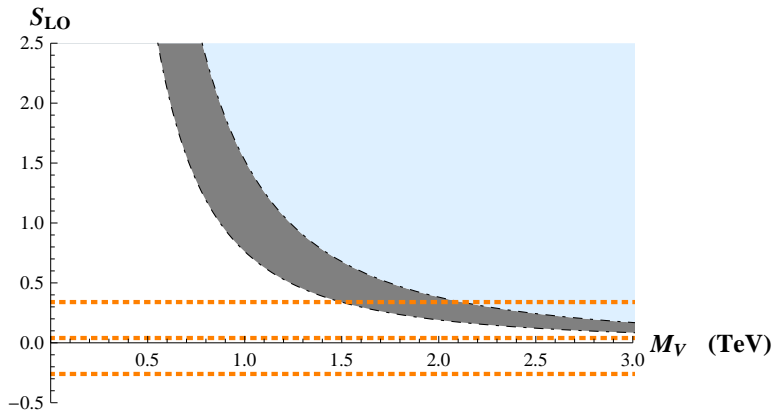}
\caption{\small
LO determination of $S$. The dark gray region assumes the two WSRs, while the allowed range gets enlarged to the light-blue region if one only assumes the first WSR and $M_A > M_V$. The horizontal dashed lines show the experimentally allowed region at $3\sigma$. 
 }
\label{fig.allowed-LO}
\end{figure}

\begin{figure}[t!]
\includegraphics[scale=1,width=7.1cm]{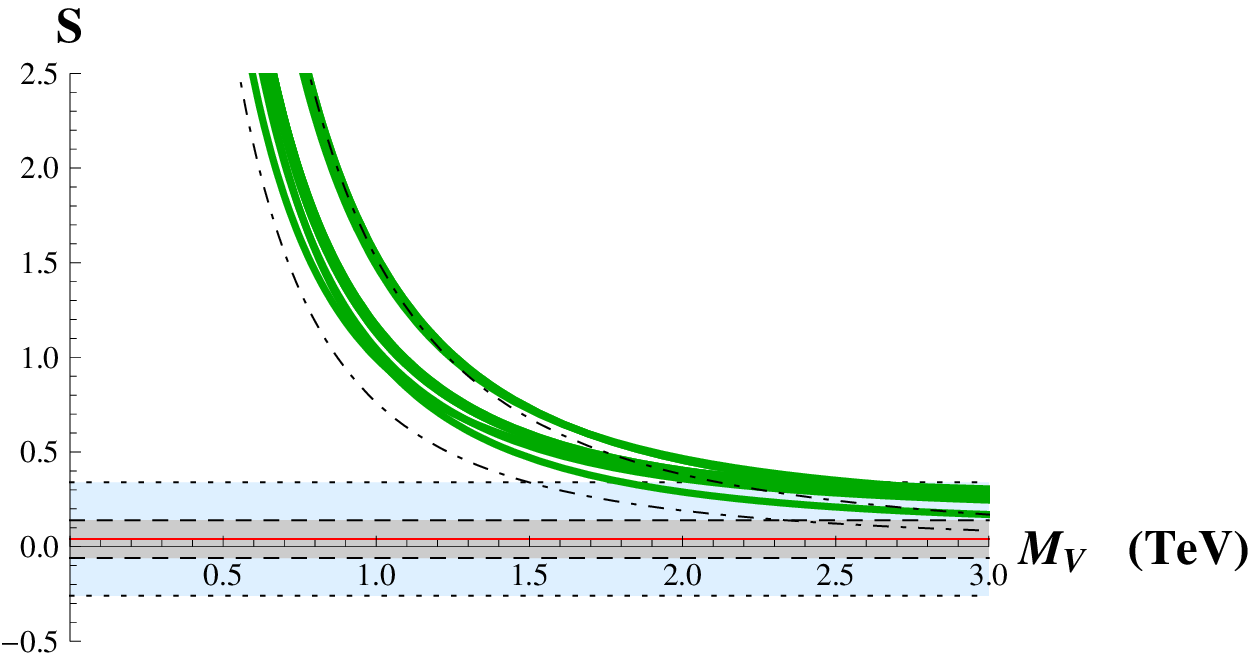}
\hskip .5cm
\includegraphics[scale=1,width=7.1cm]{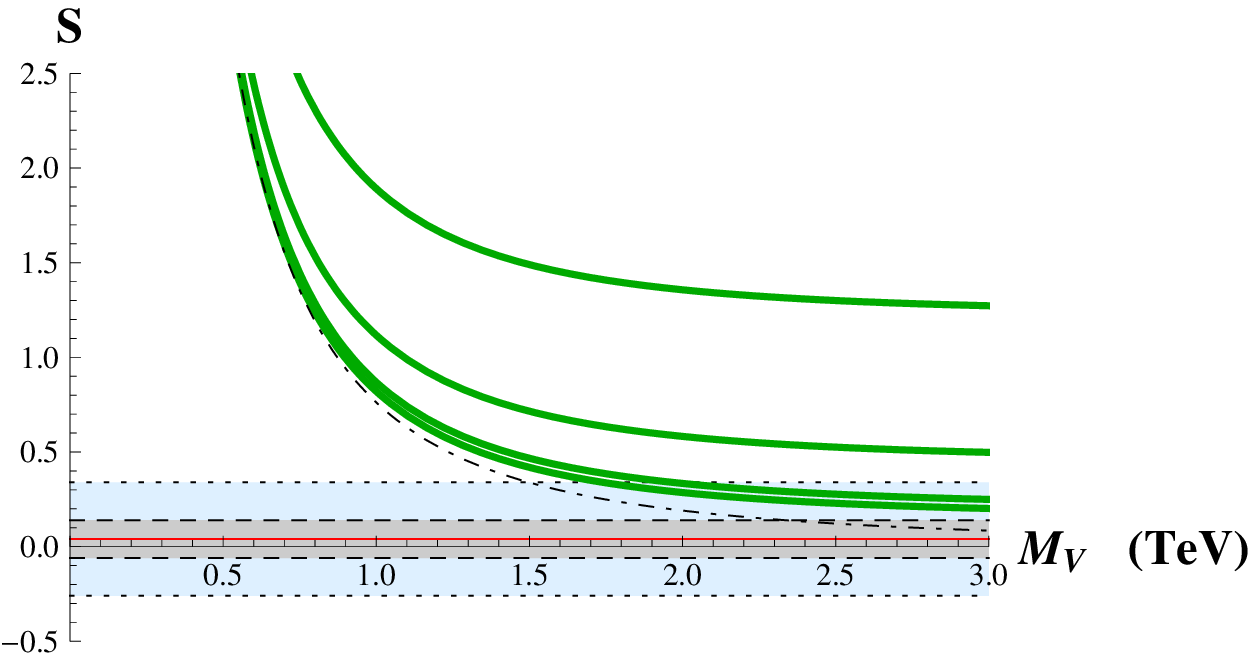}
\caption{\small NLO determination of $S$ imposing the two WSRs (left) and lower bound on $S$ imposing the first WSR plus the VFF and AFF constraints. 
The dash-dotted curves  provide the LO bounds, the horizontal dashed (dotted) lines show the experimentally allowed region at $1\sigma$ ($3\sigma$) and the red horizontal line is the experimental central value. For more details see Ref.~\cite{paper}.
}
\label{justWSR+LO-S}
\end{figure}

The global fit to precision electroweak data provides the ``experimental'' value $S = 0.04\pm 0.10$~\cite{Sparameter}, normalized to the SM reference point $M_H = 0.120$ TeV. 
\begin{enumerate}
\item {\bf LO}. Considering the first and the second  WSRs $S_{\mathrm{LO}}$ becomes $S_{\mathrm{LO}} = 4\pi v^2 M_V^{-2} \, ( 1 + M_V^2/M_A^2 )$~\cite{Peskin:92}. Since the WSRs imply $M_A>M_V$, the prediction turns out to be bounded by $4\pi v^2 /M_V^{2} < S_{\rm   LO}  <  8 \pi v^2/M_V^2$. If only the first WSR is considered, and assuming  $M_A>M_V$, one obtains for $S$ the lower bound $S_{\mathrm{LO}} = 4\pi \left\{ v^2/M_V^2+ F_A^2 \left( 1/M_V^2 - 1/M_A^2 \right) \right\} > 4\pi v^2/M_V^2$. The resonance masses need to be heavy enough to comply with the strong experimental bound, see Fig.~\ref{fig.allowed-LO}. The experimental data implies $M_V > 1.5~$TeV at the $3\sigma$ level. 
\item {\bf NLO with the 1st and the 2nd WSRs}. The four short-distance constraints on the two-particle spectral function plus the two NLO (LO) WSRs allow us to determine 6 parameters, being $M_V$ the only free parameter.
We have found eight sets of consistent solutions~\cite{paper}. The corresponding prediction for $S_{\mathrm{NLO}}$ is shown in Fig.~\ref{justWSR+LO-S}. Note that the differences with respect to the LO estimate are not very large. In order to obtain a  value of $S$ compatible with the experimental band, one needs roughly the same range of masses as at tree-level: $M_V>1.8$~TeV at the 3$\sigma$ level.

\item {\bf NLO with only the 1st WSR}. Without the second WSR we can no-longer determine $F_V^r$ and $F_A^r$. Therefore, we can only derive lower bounds on $S$~\cite{paper}.
Without imposing the second WSR and assuming $M_A>M_V$ we have found one reliable solution, being $M_V$ and the mass ratio $M_A/M_V$ unfixed. 
Fig.~\ref{justWSR+LO-S} shows the predicted lower bounds on $S$ for 
various mass ratios~\cite{paper}. Note that again one needs $M_V>1.8$~TeV to reach compatibility with  the ``experimental'' data at the $3\sigma$ level.
\end{enumerate}

\section{Conclusions}

We have presented a one-loop calculation of the oblique $S$ parameter within Higgsless models of EWSB and have analyzed the phenomenological implications of the available electroweak precision data~\cite{paper}. Strongly-coupled models of EWSB are characterized by the presence of massive resonance states. We have considered the lightest vector and axial-vector resonances. Our calculation takes advantage of the dispersive representation of $S$~\cite{Peskin:92} and the low-energy couplings are determined through short-distance conditions. 

We have found that the $S$ parameter requires a high resonance mass scale, $M_V>1.8\,$TeV, in most strongly-coupled scenarios of EWSB. It has been suggested recently that if these resonances exist, they should have masses above $1\,$TeV~\cite{pedro}.

\begin{theacknowledgments}
 We wish to thank the organizers of QCD@work 2012 for the pleasant congress and A.~Pich and J.J.Sanz-Cillero for their comments. This work has been supported in part by the Spanish Government [grants FPA2007-60323, FPA2011-23778 and CSD2007-00042 (CPAN)] and the Universidad CEU Cardenal Herrera [PRCEU-UCH35/11].

\end{theacknowledgments}

\bibliographystyle{aipproc}   

\bibliography{sample}

\begin{thebibliography}{9}

\bibitem{HK-mechanism}
F. Englert and R. Brout, Phys.\ Rev.\ Lett.\ {\bf 13} (1964) 321;\\
  P.~W.~Higgs,
  Phys.\ Rev.\  {\bf 145} (1966) 1156;
  Phys.\ Rev.\ Lett.\  {\bf 13} (1964) 508;\\
  G.~S.~Guralnik, C.~R.~Hagen and T.~W.~B.~Kibble,
  Phys.\ Rev.\ Lett.\  {\bf 13} (1964) 585;\\
T.W.B. Kibble, Phys.\ Rev.\ {\bf 155} (1967) 1554.

\bibitem{LHC}
  S.~Chatrchyan {\it et al.}  [CMS Collaboration],
  Phys.\ Lett.\ B
  [arXiv:1207.7235 [hep-ex]];\\
  G.~Aad {\it et al.}  [ATLAS Collaboration],
  Phys.\ Lett.\ B
  [arXiv:1207.7214 [hep-ex]].

\bibitem{AB:80} T. Appelquist and C. Bernard, Phys.\ Rev.\ D {\bf 22} (1980) 200.

\bibitem{EWSB}
  R.~S.~Chivukula,
  in {\it Probing The Standard Model of Particle Interactions}
  (Les Houches LXVIII, 1997), eds. R. Gupta et al. (Elsevier Sci. B.V., Amsterdam, 1999),
  Vol. II, p.~1339
  [hep-ph/9803219];\\
  A.~Pomarol,
  CERN Yellow Report CERN-2012-001, p.~115
  [arXiv:1202.1391 [hep-ph]];\\
%
  J.~R.~Andersen {\it et al.},
  Eur.\ Phys.\ J.\ Plus {\bf 126} (2011) 81
  [arXiv:1104.1255 [hep-ph]].

\bibitem{technicolor}
    S.~Weinberg,
    Phys.\ Rev.\ D {\bf 19} (1979) 1277; {\bf 13} (1976) 974; \\
    L.~Susskind,
    Phys.\ Rev.\ D {\bf 20} (1979) 2619.

\bibitem{paper}
  A.~Pich, I.~Rosell and J.~J.~Sanz-Cillero,
  JHEP {\bf 1208} (2012) 106
  [arXiv:1206.3454 [hep-ph]].

\bibitem{Peskin:92}
  M. E. Peskin and T. Takeuchi,
  Phys.\ Rev.\ D {\bf 46} (1992) 381;
  Phys.\ Rev.\ Lett.\  {\bf 65} (1990) 964.

\bibitem{RChT}
  G.~Ecker, J.~Gasser, A.~Pich and E.~de Rafael,
  Nucl.\ Phys.\ B {\bf 321} (1989) 311; \\
  G.~Ecker {\it et al.}, 
  Phys.\ Lett.\ B {\bf 223} (1989) 425; \\
  V.~Cirigliano {\it et al.},
  Nucl.\ Phys.\ B {\bf 753} (2006) 139
 [hep-ph/0603205].

\bibitem{estimating-LECs}
  O.~Cata and S.~Peris,
  Phys.\ Rev.\ D {\bf 65} (2002) 056014
  [hep-ph/0107062];\\
%
  I.~Rosell, J.~J.~Sanz-Cillero and A.~Pich,
  JHEP {\bf 0408} (2004) 042
  [hep-ph/0407240]; 
%
  JHEP {\bf 0701} (2007) 039
  [hep-ph/0610290];
%
  JHEP {\bf 1102} (2011) 109
  [arXiv:1011.5771 [hep-ph]].

\bibitem{L10}
%
  A.~Pich, I.~Rosell and J.~J.~Sanz-Cillero,
  JHEP {\bf 0807} (2008) 014
  [arXiv:0803.1567 [hep-ph]].

\bibitem{other}
  S.~Matsuzaki, R.~S.~Chivukula, E.~H.~Simmons and M.~Tanabashi,
  Phys.\ Rev.\ D {\bf 75} (2007) 073002
  [hep-ph/0607191],
 075012
  [hep-ph/0702218];\\
%
    R. Barbieri,  G. Isidori,  V.S. Rychkov and E. Trincherini,
    Phys. Rev. D {\bf 78} (2008) 036012
    [arXiv:0806.1624 [hep-ph]];\\
%
    O. Cata and J.F. Kamenik,
    Phys. Rev. D {\bf 83} (2011) 053010
    [arXiv:1010.2226 [hep-ph]];\\
%
    A. Orgogozo and  S. Rychkov,
JHEP {\bf 1203} (2012) 046    [arXiv:1111.3534 [hep-ph]].

\bibitem{WSR}
  S.~Weinberg,
  Phys.\ Rev.\ Lett.\  {\bf 18} (1967) 507.

\bibitem{Sparameter}
http://gfitter.desy.de/; 
LEP Electroweak Working Group, http://lepewwg.web.cern.ch/LEPEWWG/.

\bibitem{pedro}
  A.~Filipuzzi, J.~Portoles and P.~Ruiz-Femenia,
  JHEP {\bf 1208} (2012) 080
  [arXiv:1205.4682 [hep-ph]].

\end{thebibliography}


\end{document}